\documentclass[aps,prd,reprint,onecolumn,10pt,eqsecnum,showpacs,nofootinbib,amsfonts,amssymb,amsmath,longbibliography,superscriptaddress,floatfix]{revtex4-1}
\usepackage[colorlinks=true,allcolors=blue, pdfusetitle]{hyperref}
\usepackage{bm}
\usepackage{mathrsfs}
\usepackage[dvipsnames]{xcolor}
\usepackage{mathtools}
\usepackage{graphicx}
\usepackage{orcidlink}
\usepackage{tensor}

\usepackage{algorithm}
\usepackage{algpseudocode}

\usepackage[commentmarkup=uwave]{changes}
\definechangesauthor[name=hs, color=cyan]{hs}

\usepackage{fontawesome5}

\usepackage{ragged2e}

\usepackage{tikz}

\usetikzlibrary{shapes.geometric, arrows,positioning}

\tikzstyle{startstop} = [rectangle, rounded corners, minimum width=3cm, minimum height=1cm, text centered, draw=black, fill=orange!30]
\tikzstyle{process} = [rectangle,rounded corners, minimum width=3cm, minimum height=1cm, text centered, draw=black, fill=blue!15]
\tikzstyle{decision} = [rectangle,rounded corners, minimum width=3cm, minimum height=1cm, text centered, draw=black, fill=green!20]
\tikzstyle{arrow} = [->, >=stealth]

\graphicspath{{../plots/}}

\usepackage[T1]{fontenc}
\usepackage{lmodern}
\usepackage[utf8]{inputenc}
\usepackage{microtype}

\DeclareMathAlphabet{\mathbfsf}{\encodingdefault}{\sfdefault}{bx}{sl}

\usepackage[normalem]{ulem}

\begin{document}
	\title{Estimating the strength of Lorentzian distribution in non-commutative geometry by solar system tests}
	
	\author{Rui-Bo Wang\,\orcidlink{0009-0003-3198-3310}}
	\email[Rui-Bo Wang: ]{wangrb2021@lzu.edu.cn}
	\affiliation{School of Physical Science and Technology, Lanzhou University, Lanzhou, Gansu 730000, China}
	
	\author{Shi-Jie Ma\,\orcidlink{0000-0003-4423-0142}}
	\email[Shi-Jie Ma: ]{220220939731@lzu.edu.cn}
	\affiliation{School of Physical Science and Technology, Lanzhou University, Lanzhou, Gansu 730000, China}
	
	\author{Jian-Bo Deng\,\orcidlink{0000-0002-0586-6220}}
	\email[Jian-Bo Deng(corresponding author): ]{dengjb@lzu.edu.cn}
	\affiliation{School of Physical Science and Technology, Lanzhou University, Lanzhou, Gansu 730000, China}
	
	\author{Xian-Ru Hu\,\orcidlink{0000-0001-9818-8635}}
	\email[Xian-Ru Hu: ]{huxianru@lzu.edu.cn}
	\affiliation{School of Physical Science and Technology, Lanzhou University, Lanzhou, Gansu 730000, China}
	
	\date{\today}
	
	\begin{abstract}
		In this paper, we study four classical tests of Schwarzschild space-time with Lorentzian distribution in non-commutative geometry. We performed detailed calculations of the first-order corrections induced by the non-commutative parameter on planetary orbital precession, light deflection, radar wave delay, and gravitational redshift. The study showed that the impact of the non-commutative parameter on the time-like geodesics is significantly greater than its effect on the null geodesics. By using a series of precise experimental observations, the allowable range for the non-commutative parameter is ultimately constrained within $\Theta\leq0.067579~\mathrm{m}^{2}$, which is given by Mercury's orbital precession. This result aligns with the view that $\sqrt{\Theta}$ is of the order of the Planck length. Moreover, this constrained parameter range exceeds the Planck scale by a significant margin.

	\end{abstract}

	\maketitle
	\section{Introduction}\label{Sect1}
	Modified gravity theory is a classic and widely studied research direction in the field of general relativity (GR)~\cite{mg1,mg2,mg3,mg4,mg5,mg6,mg7,mg8,mg9,mg10,mg11}. Like quantum gravity~\cite{qg1,qg2,qg3,qg4,qg5,qg6,qg7,qg8}, non-commutative geometry is also a theory of gravitational quantization~\cite{noncomu1,noncomu2,noncomu3,noncomu4,noncomu5,noncomu6,noncomu7}. Unlike in traditional quantum mechanics, non-commutative geometry posits that space-time coordinate operators themselves are non-commutative, as described by a relation $\left[\hat{x}^{\mu},\hat{x}^{\nu}\right]=\mathrm{i}\hat{\Theta}^{\mu\nu}$, where $\hat{x}^{\mu}$ is space-time coordinate operator and $\hat{\Theta}^{\mu\nu}$ is an anti-symmetric constant matrix. The parameter $\sqrt{\Theta}$ represents the minimum scale of space-time. While the precise value of $\Theta$ remains undetermined, $\sqrt{\Theta}$ is generally considered to be of the order of the Planck length~\cite{commutator,planck-length}. There have been extensive research conducted to study the influences brought by non-commutative geometry in theories of gravity~\cite{gr1,gr2,gr3,gr4,gr5,gr6,gr7,gr8}. An especially intriguing perspective is that, due to non-commutative effects, the mass distribution of a point particle, originally concentrated at a single point, is instead modified as a continuous distribution spreading throughout the entire space. This offers a theoretical avenue for potentially addressing the black hole singularity problem. Two alternative forms of such matter distributions are commonly considered: the Lorentzian distribution $\rho=\frac{M}{\left(4\pi\Theta\right)^{\frac{3}{2}}}\exp\left(-\frac{r^{2}}{4\Theta}\right)$~\cite{lorentziandistribution} and the Gaussian distribution $\rho=\frac{M\sqrt{\Theta}}{\pi^{\frac{3}{2}}\left(r^{2}+\pi\Theta\right)^{2}}$~\cite{gaussiandistribution1,gaussiandistribution2,gaussiandistribution3}. When $\Theta$ approaches zero, these two distributions both converge to a Dirac delta function $M\delta^{3}\left(x\right)$, leading to a standard Schwarzschild space-time. There have been many studies investigating the properties of black holes in the background of non-commutative geometry~\cite{gr1,gr2,gr3,gr4,lorentziandistribution,gaussiandistribution1,gaussiandistribution2,gaussiandistribution3,Gaussianthermo,optics,thermo1,thermo2,thermo3,thermo4,Horizon-scale,gr9,gr10,gr11,gr12,gr13}. The effect of noncommutative geometry on the classical orbits of particles in a central force potential was investigated in~\cite{centralpotential}. Moreover, because the commutator of coordinate operators being a nonzero constant $\Theta$ matrix is incompatible with Lorentz symmetry~\cite{noncomu1}, non-commutative geometry may offer valuable insights into the breaking of Lorentz covariance, which has also attracted much interest of researchers.~\cite{lorentz1,lorentz2,lorentz2,lorentz3,lorentz4,lorentz5,lorentz6}.
	
	GR has undergone extensive experimental verification. Four of the most renowned tests include the precession of Mercury's orbit, light deflection, gravitational redshift, and radar wave delay. Due to the gravitational effect, the light will undergo a deflection when it passes near a massive celestial body. Additionally, considering the effect caused by the gravitational slowing down of light, the motion of light in a gravitational field will result in a time delay. And, the time delay can happen even when light travels in a radial direction. On the other hand, the variation in clock rates at different positions within a gravitational field results in a frequency shift of light waves, as they represent oscillations of the electromagnetic field. Additionally, because the time-like geodesics in Schwarzschild space-time does not satisfy the Binet equation in Newton's mechanics, a correction at the scale of the Schwarzschild radius emerges. Consequently, the trajectory of a planet in Schwarzschild space-time is no longer a closed elliptical orbit, with its semi-major axis shifting periodically. Although $\sqrt{\Theta}$ is considered to be on the order of the Planck length $\ell_{p} (\simeq 1.616\times10^{-35}\mathrm{m})$, it is still crucial to investigate the limit of the non-commutative parameter. In~\cite{BTZgravastar}, the authors study the relationship between the noncommutative parameter and the cosmological constant by constructing a BTZ gravastar in non-commutative geometry, and obtain a relatively precise parameter constraint.
	
	In this work, we study the parameter constraints of Lorentzian distribution in non-commutative geometry using solar system tests. Investigating the constraints on this parameter provided by the classic experiments in the Solar System is of meaningful interest. Moreover, calculating the four classical tests of GR within the framework of
	non-commutative geometry is valuable for studying the impact of non-commutative
	geometry on the motion of time-like particles and light rays. This article is organized as follows. In Sect.~\ref{Sect2}, we substitute the Lorentzian distribution into the Einstein equation to obtain the metric of Schwarzschild space-time in non-commutative geometry. In Sect.~\ref{Sect3}, we conduct detailed calculations of the four classical experiments in Schwarzschild space-time with the Lorentzian distribution, including planetary orbital precession, light deflection, radar wave delay, and gravitational redshift effects. With the help of a series of precise observational data, we determine a range of the non-commutative parameter. Finally, we give our conclusion and outlook in Sect.~\ref{Sect4}.

	\section{Schwarzschild space-time of Lorentzian distribution in non-commutative geometry}\label{Sect2} 
	We start from the Einstein equation
	\begin{equation}\label{eq2_1}
		R_{\mu}^{\nu}-\frac{1}{2}\delta_{\mu}^{\nu}R=8\pi T_{\mu}^{\nu},
	\end{equation}
	where $R_{\mu}^{\nu}$ is Ricci tensor, $\delta_{\mu}^{\nu}$ is Kronecker symbol, $R$ is Ricci scalar and $T_{\mu}^{\nu}$ is energy momentum
	tensor. 
	
	The Lorentzian distribution in non-commutative geometry is~\cite{optics,lorentziandistribution,Gaussianthermo,gr1}
	\begin{equation}\label{eq2_2}
		T_{0}^{0}=-\frac{M\sqrt{\Theta}}{\pi^{\frac{3}{2}}\left(r^{2}+\pi\Theta\right)^{2}},
	\end{equation}
	where $M$ is the mass of the central celestial body and $\Theta$ is the non-commutative parameter with dimension of $\left[\mathrm{L}^{2}\right]$. One could verify that
	\begin{equation}\label{eq2_3}
		\lim_{\Theta\rightarrow 0}T_{0}^{0}=-M\delta^{3}\left(r\right),
	\end{equation}
	where $\delta^{3}\left(r\right)$ is the three-dimensional Dirac delta function.
	
	A static spherically symmetric space-time is
	\begin{equation}\label{eq2_4}
		\mathrm{d}s^{2}=-f\left(r\right)\mathrm{d}t^{2}+\frac{1}{f\left(r\right)}\mathrm{d}r^{2}+r^{2}\mathrm{d}\theta^{2}+r^{2}\sin^{2}\theta\mathrm{d}\phi^{2},
	\end{equation}
	where $f\left(r\right)=1-\frac{2m\left(r\right)}{r}$ and $m\left(r\right)$ is the mass function, which could be derived from Eq.~\ref{eq2_1} as
	\begin{equation}\label{eq2_5}
		m\left(r\right)=-\int_{0}^{r} 4\pi r^{2} T_{0}^{0}\mathrm{d}r.
	\end{equation}
	By substituting the Lorentzian distribution, the Schwarzschild space-time in non-commutative geometry is~\cite{optics,metric1,metric2}
	\begin{equation}\label{eq2_6}
		f\left(r\right)=1-\frac{2M}{r}+\frac{8M\sqrt{\Theta}}{\sqrt{\pi}r^{2}}+\mathcal{O}\left(\Theta^\frac{3}{2}\right).
	\end{equation}
	It is evident that the metric asymptotically approaches the Schwarzschild geometry at infinity. As seen, compared to the usual Schwarzschild space-time, the Lorentzian distribution provides an additional $r^{-2}$ potential to the space-time. It is worth mentioning that in~\cite{r-2}, the authors calculated the perihelion and anomaly precession caused by various modified gravity theories (including $r^{-k}$ extra potentials). Furthermore, this book also calculates the impact of modified gravity on astronomical observables and provides valuable insights for the practical experimental detection of modified gravity theories. Moreover, in~\cite{genRN}, the authors extended the analysis in Reissner-Nordstr$\ddot{\mathrm{o}}$m (RN) space-time to the general $r^{-2}$ extra potential in space-time. These work provide highly valuable tools for studying the effect of extra potentials in modified gravity on the geodesic motion of particles.
	
	Specifically, when $r$ goes to zero, one have
	\begin{equation}\label{eq2_7}
		f\left(r\right)=1-\frac{8Mr^{2}}{3\pi^{\frac{5}{2}}\Theta^{\frac{3}{2}}}+\mathcal{O}\left(r^{4}\right).
	\end{equation}
	
	In this case, the Schwarzschild space-time in non-commutative geometry contains a de Sitter core with an effective cosmological constant,
	\begin{equation}\label{eq2_8}
		\Lambda_{\mathrm{eff}}=\frac{8M}{\pi^{\frac{5}{2}}\Theta^{\frac{3}{2}}},
	\end{equation}
	which avoids the formation of a singularity. However, since the non-commutative parameter $\Theta$ is particularly small, all the scenarios discussed in this paper correspond to $r\gg\sqrt{\Theta}$. Therefore, the use of Eq.~\ref{eq2_6} is considered to be very safe.
	
	To simplify the calculations, one could introduce a parameter with dimension of $\left[\mathrm{L}\right]$:
	\begin{equation}\label{eq2_9}
		a=\frac{8\sqrt{\Theta}}{\sqrt{\pi}},
	\end{equation}
	which also could be regarded as a parameter indicating the intensity of non-commutative geometry. The metric is expressed as
	\begin{equation}\label{eq2_10}
		f\left(r\right)=1-\frac{2M}{r}+\frac{aM}{r^{2}}.
	\end{equation}

	\section{Parameter constraints given by the solar system tests}\label{Sect3}
	In this section, we separately calculate the perihelion precession of Mercury, light deflection, time delay of light, and gravitational redshift in Schwarzschild space-time with the Lorentzian distribution. Additionally, we choose some of the more precise observational data to derive an acceptable range for the non-commutative parameter. In this section, the International System of Units (SI) is used. The metric in this case is
	\begin{equation}\label{eq3_0_1}
		f\left(r\right)=1-\frac{2GM}{c^{2}r}+\frac{aGM}{c^{2}r^{2}}.
	\end{equation}

	\subsection{Perihelion precession of planet}\label{Sect3_1}
	Considering a particle moving in the gravitational field, its geodesic equation is
	\begin{equation}\label{eq3_1_1}
		\frac{\mathrm{d}^{2}x^{\lambda}}{\mathrm{d}s^{2}}+\Gamma_{\mu\nu}^{\lambda}\frac{\mathrm{d}x^{\mu}}{\mathrm{d}s}\frac{\mathrm{d}x^{\nu}}{\mathrm{d}s}=0,
	\end{equation}
	where
	\begin{equation}\label{eq3_1_2}
		\Gamma_{\mu\nu}^{\lambda}=\frac{1}{2}g^{\lambda\sigma}\left(\partial_{\mu}g_{\sigma\nu}+\partial_{\nu}g_{\sigma\mu}-\partial_{\sigma}g_{\mu\nu}\right)
	\end{equation}
	is connection.
	
	Eq.~\ref{eq3_1_1} gives
	\begin{equation}\label{eq3_1_3}
		\frac{\mathrm{d}^{2}r}{\mathrm{d}s^{2}}-\frac{f'}{2f}\left(\frac{\mathrm{d}r}{\mathrm{d}s}\right)^{2}-rf\left(\frac{\mathrm{d}\theta}{\mathrm{d}s}\right)^{2}-rf\sin^{2}\theta \left(\frac{\mathrm{d}\phi}{\mathrm{d}s}\right)^{2}+\frac{1}{2}c^{2}ff'\left(\frac{\mathrm{d}t}{\mathrm{d}s}\right)^{2}=0,
	\end{equation}
	\begin{equation}\label{eq3_1_4}
		\frac{\mathrm{d}^{2}\phi}{\mathrm{d}s^{2}}+\frac{2}{r}\frac{\mathrm{d}r}{\mathrm{d}s}\frac{\mathrm{d}\phi}{\mathrm{d}s}+2\cot\theta\frac{\mathrm{d}\theta}{\mathrm{d}s}\frac{\mathrm{d}\phi}{\mathrm{d}s}=0.	
	\end{equation} 
	Symbol ``~$'$~" represents the derivative with respect to $r$. Without loss of generality, we set $\theta=\pi/2$ and then Eqs.~\ref{eq3_1_3} and \ref{eq3_1_4} could be rewritten as
	\begin{equation}\label{eq3_1_5}
		\frac{\mathrm{d}^{2}r}{\mathrm{d}s^{2}}-\frac{f'}{2f}\left(\frac{\mathrm{d}r}{\mathrm{d}s}\right)^{2}-rf \left(\frac{\mathrm{d}\phi}{\mathrm{d}s}\right)^{2}+\frac{1}{2}c^{2}ff'\left(\frac{\mathrm{d}t}{\mathrm{d}s}\right)^{2}=0,
	\end{equation}
	\begin{equation}\label{eq3_1_6}
		r^{2}\frac{\mathrm{d}\phi}{\mathrm{d}s}=\frac{h}{c},
	\end{equation} 
	where $h$ is a constant.
	
	For a time-like particle, its line element $\mathrm{d}s^{2}\left(<0\right)$ is
	\begin{equation}\label{eq3_1_7}
		\mathrm{d}s^{2}=-c^{2}f\mathrm{d}t^{2}+f^{-1}\mathrm{d}r^{2}+r^{2}\mathrm{d}\phi^{2},
	\end{equation}
	which results in this relation
	\begin{equation}\label{eq3_1_8}
		\left(\frac{\mathrm{d}r}{\mathrm{d}s}\right)^{2}=-f+c^{2}f^{2}\left(\frac{\mathrm{d}t}{\mathrm{d}s}\right)^{2}-r^{2}f\left(\frac{\mathrm{d}\phi}{\mathrm{d}s}\right)^{2}.
	\end{equation}
	Using Eqs.~\ref{eq3_1_3}, \ref{eq3_1_6} and \ref{eq3_1_8}, one could get
	\begin{equation}\label{eq3_1_9}
		\frac{\mathrm{d}^{2}r}{\mathrm{d}s^{2}}+\frac{f'}{2}+\frac{h^{2}f'}{2c^{2}r^{2}}-\frac{h^{2}f}{c^{2}r^{3}}=0.
	\end{equation}
	By introducing transformation $r=u^{-1}$, the above function becomes
	\begin{equation}\label{eq3_1_10}
		\frac{\mathrm{d}^{2}u}{\mathrm{d}\phi^{2}}+uf-\frac{f'}{2}-\frac{c^{2}f'}{2h^{2}u^{2}}=0.
	\end{equation}
	By substituting the metric $f(r)$, one could get the orbit equation of particles in non-commutative Schwarzschild space-time as follows
	\begin{equation}\label{eq3_1_11}
		\frac{\mathrm{d}^{2}u}{\mathrm{d}\phi^{2}}+u=\frac{GM}{h^{2}}+\frac{3GM}{c^{2}}u^{2}-\frac{aGM}{h^{2}}u-\frac{2aGM}{c^{2}}u^{3}.
	\end{equation}  
	Please notice that the Binet equation, which works in Newton's mechanics, is
	\begin{equation}\label{eq3_1_12}
		\frac{\mathrm{d}^{2}u}{\mathrm{d}\phi^{2}}+u=\frac{GM}{h^{2}},
	\end{equation}
	whose solution is a conic curve
	\begin{equation}\label{eq3_1_13}
		u_{0}=\frac{1+e\cos\phi}{p},
	\end{equation}
	where $p=\frac{h^{2}}{GM}$ and $e$ is the orbit's eccentricity. Assuming the solution of Eq.~\ref{eq3_1_11} is $u=u_{0}+r_{\mathrm{S}}u_{1}$ ($r_{\mathrm{S}}=\frac{2GM}{c^{2}}$ is Schwarzschild radius), one could derive
	\begin{equation}\label{eq3_1_14}
		r_{\mathrm{S}}\frac{\mathrm{d}^{2}u_{1}}{\mathrm{d}\phi^{2}}+r_{\mathrm{S}}u_{1}=-\frac{a}{p}u+\frac{3r_{\mathrm{S}}}{2}u^{2}-ar_{\mathrm{S}}u^{3}.
	\end{equation}
	Expand above function further:
	\begin{equation}\label{eq3_1_15}
		r_{\mathrm{S}}\frac{\mathrm{d}^{2}u_{1}}{\mathrm{d}\phi^{2}}+r_{\mathrm{S}}u_{1}=-\frac{a}{p}u_{0}+\frac{3r_{\mathrm{S}}}{2}u_{0}^{2}+\mathcal{O}(r_{\mathrm{S}}^{2}, ar_{\mathrm{S}}).
	\end{equation}
	Considering that $a$ and $r_{\mathrm{S}}$ are small, one can keep above equation to the first-order small quantity $\mathcal{O}\left(a, r_{\mathrm{S}}\right)$ and obtain
	\begin{equation}\label{eq3_1_16}
		\frac{\mathrm{d}^{2}u_{1}}{\mathrm{d}\phi^{2}}+u_{1}=\frac{\left(1+e\cos\phi\right)\left(3r_{\mathrm{S}}+3er_{\mathrm{S}}\cos\phi-2a\right)}{2p^{2}r_{\mathrm{S}}}.
	\end{equation}
	Substituting $u_{1}=c_{1}+c_{2}\phi\sin\phi+c_{3}\cos2\phi$, one could obtain the undetermined coefficients:
	\begin{equation}\label{eq3_1_17}
		c_{1}=\frac{6r_{\mathrm{S}}+3e^{2}r_{\mathrm{S}}-4a}{4p^{2}r_{\mathrm{S}}},~~
		c_{2}=\frac{e\left(3r_{\mathrm{S}}-a\right)}{2p^{2}r_{\mathrm{S}}},~~
		c_{3}=-\frac{e^{2}}{4p^{2}}.
	\end{equation}
	Now the equation of motion for a particle is
	\begin{equation}\label{eq3_1_18}
		u=\frac{1}{p}\left(1+e\cos\phi+\frac{6r_{\mathrm{S}}+3e^{2}r_{\mathrm{S}}-4a}{4p}+\frac{e\left(3r_{\mathrm{S}}-a\right)}{2p}\phi\sin\phi-\frac{r_{\mathrm{S}}e^{2}}{4p}\cos2\phi\right).
	\end{equation}
	Now we study the precession. In above expression, the perturbation of constant term will only cause a small deviation from the original orbit $u_{0}$ and the periodic term $\cos2\phi$ doesn't make a long-term correction of the orbit. Only the term $\phi\sin\phi$ plays a significant role as $\phi$ increases. So one could only consider this effective orbit equation
	\begin{equation}\label{eq3_1_19}
		u=\frac{1}{p}\left(1+e\cos\phi+e\epsilon\phi\sin\phi\right),
	\end{equation}
	where $\epsilon=\frac{3r_{\mathrm{S}}-a}{2p}$ is a small quantity. One could write
	\begin{equation}\label{eq3_1_20}
		\cos\left(\phi-\epsilon\phi\right)=\cos\phi\cos\epsilon\phi+\sin\phi\sin\epsilon\phi=\cos\phi+\epsilon\phi\sin\phi.
	\end{equation}
	So Eq.~\ref{eq3_1_19} could be written as
	\begin{equation}\label{eq3_1_21}
		u=\frac{1}{p}\left(1+e\cos\Phi\right),
	\end{equation}
	where $\Phi=\phi\left(1-\epsilon\right)$. Planet's aphelion meets
	\begin{equation}\label{eq3_1_22}
		\Phi_{n}=\left(2n+1\right)\pi,~~n\in\mathbb{Z},
	\end{equation}
	so planet's angular increase for one period is
	\begin{equation}\label{eq3_1_23}
		\phi_{n+1}-\phi_{n}=\frac{\Phi_{n+1}-\Phi_{n}}{1-\epsilon}=2\pi\left(1+\epsilon\right),
	\end{equation}
	which gives the precession angle for one period
	\begin{equation}\label{eq3_1_24}
		\Delta=2\pi\epsilon=\frac{6\pi GM}{c^{2}p}-\frac{a\pi}{p}.
	\end{equation}
	\begin{figure}[htbp]
		\centering
		\includegraphics[width=0.8\textwidth]{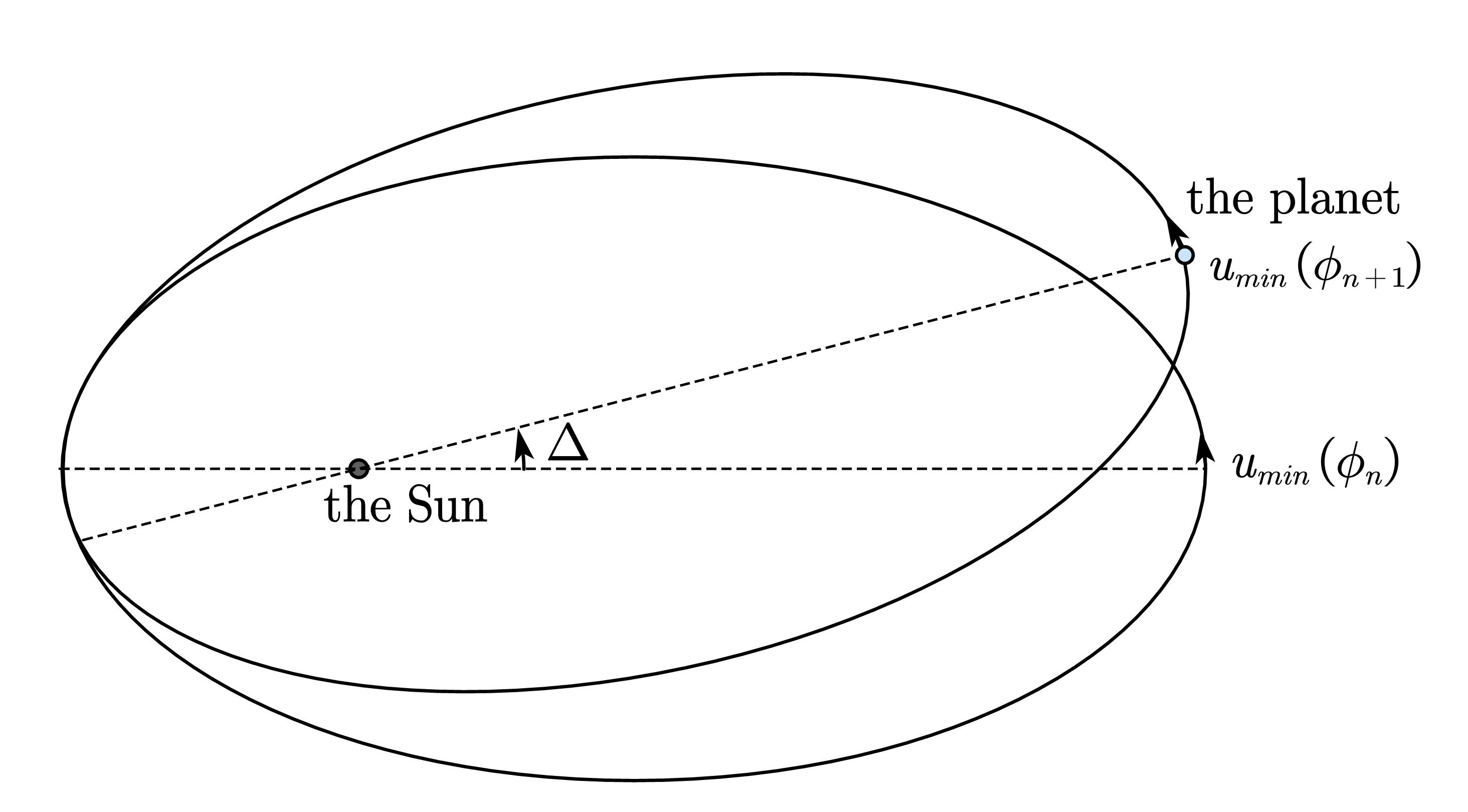}
		\caption{The precession of planetary orbits.}\label{precession}
	\end{figure}
	It is clear that $\frac{6\pi GM}{c^{2}p}$ is exactly the prediction in GR. Fig.~\ref{precession} is a schematic diagram illustrating the precession of planetary orbits.
	
	According to Kepler's third law
	\begin{equation}\label{eq3_1_25}
		\frac{R^{3}}{T^{2}}=\frac{GM}{4\pi^{2}},
	\end{equation}
	where $R=\frac{p}{1-e^{2}}$ is semi-major axis of elliptic orbit, $T$ is period. Then Eq.~\ref{eq3_1_24} reads
	\begin{equation}\label{eq3_1_26}
		\Delta=\frac{24\pi^{3}R^{2}}{c^{2}T^{2}\left(1-e^{2}\right)}-\frac{a\pi}{R\left(1-e^{2}\right)}
	\end{equation}
	The precession angle for one century is
	\begin{equation}\label{eq3_1_27}
		\Delta_{\mathrm{c}}=3.83772\frac{R^{2}}{T^{3}\left(1-e^{2}\right)}-4.33161\times 10^{-4}\frac{a}{RT\left(1-e^{2}\right)}.
	\end{equation}
	It is worth noticing that in above formula, the precession angle $\Delta_{\mathrm{c}}$ is measured in arcseconds $''$, the semi-major axis $R$ is measured in astronomical units AU (1AU=149597870700 m), the period $T$ is measured in sidereal years (1~yr=365.256 days), and the non-commutative parameter $a$ is measured in meters m. For Mercury, its semi-major axis, orbital eccentricity, and orbit period are $R=0.38709893$ AU, $e=0.20563069$ annd $T=87.969$ days respectively~\cite{nasadata}. Regarding the experimental tests of gravitational theory within the solar system, it is worth mentioning E.V. Pitjeva, A. Fienga, and L. Iorio. In recent years, each of their teams has conducted a considerable amount of valuable work on this subject~\cite{exp1,exp2,exp3,exp4,exp5,exp6,exp7,exp8,exp9,exp10,exp11}. In this paper, we just take the data from~\cite{Mercury} as an example to estimate the upper limit of $\Theta$. Mercury's precession was reported as $\Delta_{\mathrm{c}}=\left(42.979\pm0.003\right)''\mathrm{Century}^{-1}$~\cite{Mercury}. It gives this following constraint
	\begin{equation}\label{eq3_1_28}
		-0.063449~\mathrm{m}\leq a \leq 1.1733~\mathrm{m}.
	\end{equation}
	It is clear that according to Eq.~\ref{eq2_9}, $a$ must be positive, but we still keep the negative result to provide a reference for the possible theoretical study in the future. With relation $a=\frac{8\sqrt{\Theta}}{\sqrt{\pi}}$, the constraint of $\Theta$ is
	\begin{equation}\label{eq3_1_29}
		\Theta\leq0.067579~\mathrm{m}^{2} 		
	\end{equation}

	\subsection{Deflection of light}\label{Sect3_2}
	Now we continue to study the deflection of light. One could only replace the parameter $s$ in Eqs.~\ref{eq3_1_5} and \ref{eq3_1_6} with the affine parameter $\lambda$ to obtain the equation of motion for light
	\begin{equation}\label{eq3_2_1}
		\frac{\mathrm{d}^{2}r}{\mathrm{d}\lambda^{2}}-\frac{f'}{2f}\left(\frac{\mathrm{d}r}{\mathrm{d}\lambda}\right)^{2}-rf \left(\frac{\mathrm{d}\phi}{\mathrm{d}\lambda}\right)^{2}+\frac{1}{2}c^{2}ff'\left(\frac{\mathrm{d}t}{\mathrm{d}\lambda}\right)^{2}=0,
	\end{equation}
	\begin{equation}\label{eq3_2_2}
		r^{2}\frac{\mathrm{d}\phi}{\mathrm{d}\lambda}=k,
	\end{equation} 
	where $k$ is a constant. And for photon, its line element is zero
	\begin{equation}\label{eq3_2_3}
		\mathrm{d}s^{2}=-c^{2}f\mathrm{d}t^{2}+f^{-1}\mathrm{d}r^{2}+r^{2}\mathrm{d}\phi^{2}=0,
	\end{equation}
	which gives
	\begin{equation}\label{eq3_2_4}
		\left(\frac{\mathrm{d}r}{\mathrm{d}\lambda}\right)^{2}=c^{2}f^{2}\left(\frac{\mathrm{d}t}{\mathrm{d}\lambda}\right)^{2}-r^{2}f\left(\frac{\mathrm{d}\phi}{\mathrm{d}\lambda}\right)^{2}.
	\end{equation}
	Similar to the derivation of planetary orbit precession, using Eqs.~\ref{eq3_2_1}, \ref{eq3_2_2} and \ref{eq3_2_4}, with the transform $r=u^{-1}$, one could obtain the equation of photon's orbit
	\begin{equation}\label{eq3_2_5}
		\frac{\mathrm{d}^{2}u}{\mathrm{d}\phi^{2}}+u=\frac{3r_{\mathrm{S}}}{2}u^{2}-ar_{\mathrm{S}}u^{3}.
	\end{equation}
	Assume $u_{0}$ meets with equation
	\begin{equation}\label{eq3_2_6}
		\frac{\mathrm{d}^{2}u_{0}}{\mathrm{d}\phi^{2}}+u_{0}=0,
	\end{equation}
	whose solution is a straight line
	\begin{equation}\label{eq3_2_7}
		u_{0}=\frac{\sin\phi}{b},
	\end{equation}
	which represents the motion of photon without gravitational field. Assuming that the solution of Eq.~\ref{eq3_2_5} is $u=u_{0}+r_{\mathrm{S}}u_{1}$, one could write
	\begin{equation}\label{eq3_2_8}
		r_{\mathrm{S}}\frac{\mathrm{d}^{2}u_{1}}{\mathrm{d}\phi^{2}}+r_{\mathrm{S}}u_{1}=\frac{3r_{\mathrm{S}}}{2}u_{0}^{2}-ar_{\mathrm{S}}u_{0}^{3}+\mathcal{O}\left(r_{\mathrm{S}}^{2}\right).
	\end{equation}
	Neglecting the higher-order terms of $r_{\mathrm{S}}$, then $u_{1}$ should satisfy
	\begin{equation}\label{eq3_2_9}
		\frac{\mathrm{d}^{2}u_{1}}{\mathrm{d}\phi^{2}}+u_{1}=\frac{3\sin^{2}\phi}{2b^{2}}-\frac{a\sin^{3}\phi}{b^{3}}.
	\end{equation}
	The solution of this equation has this form \begin{equation}\label{eq3_2_10}
		u_{1}=c_{1}+c_{2}\cos\phi+c_{3}\phi\sin\phi+c_{4}\cos2\phi+c_{5}\sin3\phi.
	\end{equation}
	Substituting this into Eq.~\ref{eq3_2_9}, one could work out the coefficients
	\begin{equation}\label{eq22_11}
		c_{1}=\frac{3}{4b^{2}},~~
		c_{3}=\frac{3a}{8b^{3}},~~
		c_{4}=\frac{1}{4b^{2}},~~
		c_{5}=-\frac{a}{32b^{3}}.
	\end{equation}
	Without loss of generality, we set $u'\left(\frac{\pi}{2}\right)=0$, which means photon reaches its perihelion at $\phi=\frac{\pi}{2}$. It gives the last undetermined coefficient $c_{2}=-\frac{3a\pi}{16b^{3}}$.
	
	Now photon's orbit is finally obtained
	\begin{equation}\label{eq3_2_12}
		u=\frac{\sin\phi}{b}+r_{\mathrm{S}}\left(\frac{3}{4b^{2}}-\frac{3a\pi}{16b^{3}}\cos\phi+\frac{3a}{8b^{3}}\phi\sin\phi+\frac{1}{4b^{2}}\cos2\phi-\frac{a}{32b^{3}}\sin3\phi\right).
	\end{equation}
	Considering that the photon's deflection angle is small, above formula could be approximately written as
	\begin{equation}\label{eq3_2_13}
		u=\left(\frac{1}{b^{2}}-\frac{3a\pi}{16b^{3}}\right)r_{\mathrm{S}}+\left(\frac{1}{b}+\frac{9ar_{\mathrm{S}}}{32b^{3}}\right)\phi+\mathcal{O}\left(\phi^{2}\right).
	\end{equation}
	When photon is at infinity, $u=0$ should be satisfied. One could solve $\phi$
	\begin{equation}\label{eq3_2_14}
		\phi=\frac{2r_{\mathrm{S}}\left(3a\pi-16b\right)}{32b^{2}+9ar_{\mathrm{S}}}=\frac{\left(3a\pi-16b\right)r_{\mathrm{S}}}{16b^{2}}+\mathcal{O}\left(r_{\mathrm{S}}^{2}\right).
	\end{equation}
	The total deflection of light is
	\begin{equation}\label{eq3_2_15}
		\Delta\phi=-2\phi=\frac{4GM}{c^{2}b}-\frac{3aGM\pi}{4c^{2}b^{2}}.
	\end{equation}
	It could be found that term $\frac{4GM}{c^{2}b}$ is the theoretical result in GR. 
	\begin{figure}[htbp]
		\centering
		\includegraphics[width=1\textwidth]{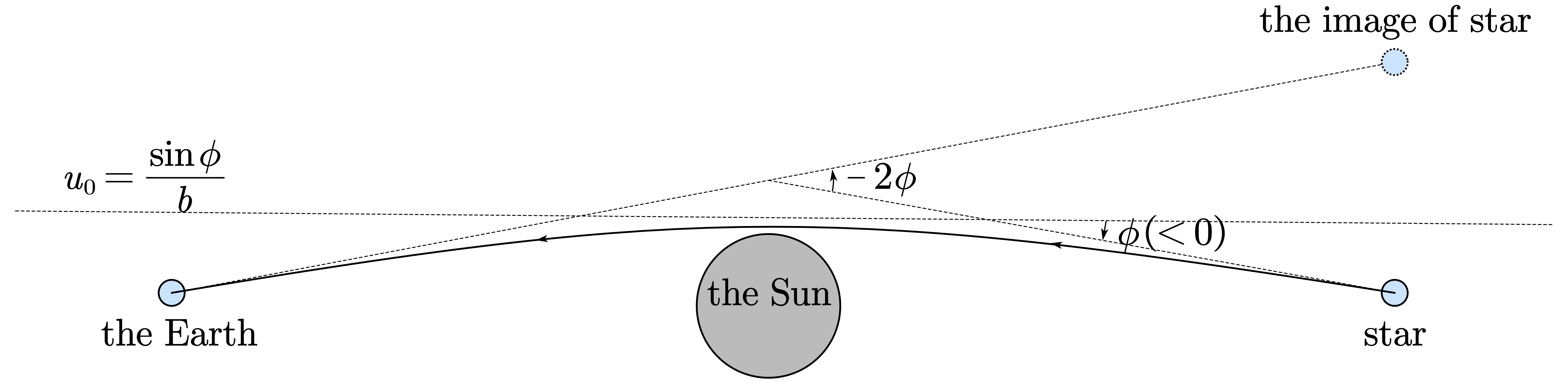}
		\caption{The bending of light near the Sun.}\label{bending}
	\end{figure}
	Fig.~\ref{bending} is a schematic diagram illustrating the deflection of light under the gravitational field of the Sun.
	
	For light just grazing the edge of the Sun, one have $M=M_{\odot}$ and $ b=R_{\odot}$.
	
	The observed value could be written as
	\begin{equation}\label{eq3_2_16}
		\Delta\phi=\frac{2\left(1+\gamma\right)GM_{\odot}}{c^{2}R_{\odot}},
	\end{equation}
	where $\gamma$ is the post-Newtonian parameter, which is measured as $\gamma=0.99992\pm0.00012$~\cite{bending}. By substituting the data of the Sun~\cite{nasadata}, one could obtain this parameter constraint
	\begin{equation}\label{eq3_2_17}
		-2.3621\times10^{4}~\mathrm{m}\leq a \leq1.1811\times10^{5}~\mathrm{m}
	\end{equation}
	or
	\begin{equation}\label{eq3_2_18}
		\Theta\leq6.8481\times10^{8}~\mathrm{m^{2}}.
	\end{equation}

	\subsection{Time delay of radar echo}\label{Sect3_3}
	To study the time delay of light in gravitational field, we use the method illustrated in~\cite{RNtime}. In this reference, the post-post-Newtonian amendment is applied to calculate the time delay in RN space-time. The inclusion of $\mathcal{O}\left(c^{-4}\right)$ is necessary for calculating the time delay in RN space-time because the electric charge $Q$ only becomes significant at this magnitude. In this paper, the post-Newtonian approximation is chosen because the non-commutative parameter $a$ already becomes obvious when considering the magnitude of $\mathcal{O}\left(c^{-2}\right)$.
	
	The fourth component of the null geodesic Eq.~\ref{eq3_1_1} (replacing $s$ with the affine parameter $\lambda$) gives
	\begin{equation}\label{eq3_3_1}
		\frac{\mathrm{d}^{2}t}{\mathrm{d}\lambda^{2}}+\frac{f'}{f}\frac{\mathrm{d}r}{\mathrm{d}\lambda}\frac{\mathrm{d}t}{\mathrm{d}\lambda}=0,
	\end{equation}
	which causes another conserved quantity $E$
	\begin{equation}\label{eq3_3_2}
		f\frac{\mathrm{d}t}{\mathrm{d}\lambda}=\frac{E}{c^{2}}.
	\end{equation}
	Using Eqs.~\ref{eq3_2_2}, \ref{eq3_2_4} and \ref{eq3_3_2}, one get
	\begin{equation}\label{eq3_3_3}
		\left(\frac{\mathrm{d}r}{\mathrm{d}t}\right)^{2}=c^{2}f^{2}-\frac{k^{2}}{E^{2}}\frac{c^{4}f^{3}}{r^{2}}.
	\end{equation}
	Photon's perihelion $r_{0}$ meets with $\frac{\mathrm{d}r}{\mathrm{d}\lambda}=0$, so Eq.~\ref{eq3_2_4} gives
	\begin{equation}\label{eq3_3_4}
		\frac{k^{2}}{E^{2}}=\frac{c^{2}f\left(r_{0}\right)}{r_{0}^{2}}.
	\end{equation}
	Using Eqs.~\ref{eq3_3_3} and \ref{eq3_3_4}, one could derive
	\begin{equation}\label{eq3_3_5}
		\frac{\mathrm{d}t}{\mathrm{d}r}=\frac{1}{c}\frac{1}{f\left(r\right)}\left(1-\frac{f\left(r\right)}{f\left(r_{0}\right)}\frac{r_{0}^{2}}{r^{2}}\right)^{-\frac{1}{2}}.
	\end{equation}
	By substituting the metric into this formula and keeping the result up to $\mathcal{O}\left(c^{-2}\right)$:
	\begin{equation}\label{eq3_3_6}
		1-\frac{f\left(r\right)}{f\left(r_{0}\right)}\frac{r_{0}^{2}}{r^{2}}=\left(1-\frac{r_{0}^{2}}{r^{2}}\right)\left(1+\frac{GM\left(ar+ar_{0}-2rr_{0}\right)}{c^{2}r^{2}\left(r+r_{0}\right)}\right)+\mathcal{O}\left(c^{-4}\right),
	\end{equation}
	one have
	\begin{equation}\label{eq3_3_7}
		\frac{1}{f\left(r\right)}\left(1-\frac{f\left(r\right)}{f\left(r_{0}\right)}\frac{r_{0}^{2}}{r^{2}}\right)^{-\frac{1}{2}}=\left(1-\frac{r_{0}^{2}}{r^{2}}\right)^{-\frac{1}{2}}\left(1+\frac{2GM}{c^{2}r}+\frac{GMr_{0}}{c^{2}r\left(r+r_{0}\right)}-\frac{3aGM}{2c^{2}r^{2}}\right)+\mathcal{O}\left(c^{-4}\right).
	\end{equation}
	Therefore, the time (coordinate time $t$) required for a photon to move from its perihelion $r_{0}$ to a point $r$ is
	\begin{equation}\label{eq3_3_8}
		\begin{aligned}
			t\left(r_{0}\rightarrow r\right)=&\frac{1}{c}\int_{r_{0}}^{r}\left(1-\frac{r_{0}^{2}}{r^{2}}\right)^{-\frac{1}{2}}\left(1+\frac{2GM}{c^{2}r}+\frac{GMr_{0}}{c^{2}r\left(r+r_{0}\right)}-\frac{3aGM}{2c^{2}r^{2}}\right)\mathrm{d}r\\
			=&\frac{\sqrt{r^{2}-r_{0}^{2}}}{c}+\frac{GM}{c^{3}}\sqrt{\frac{r-r_{0}}{r+r_{0}}}+\frac{2GM}{c^{3}}\ln\frac{r+\sqrt{r^{2}-r_{0}^{2}}}{r_{0}}\\
			&-\frac{3aGM\pi}{4c^{3}r_{0}}+\frac{3aGM}{c^{3}r_{0}}\arctan \frac{r-\sqrt{r^{2}-r_{0}^{2}}}{r_{0}}.
		\end{aligned}
	\end{equation}
	When $r \gg r_{0}$, above formula could be approximated as
	\begin{equation}\label{eq3_3_9}
		t\left(r_{0}\rightarrow r\right)=\frac{r}{c}+\frac{2GM}{c^{3}}\ln\frac{2r}{r_{0}}+\frac{GM}{c^{3}}-\frac{3aGM\pi}{4c^{3}r_{0}}.
	\end{equation}
	Assuming that a radar wave emits from the Earth, traveling through the Sun's gravitational field, then reaches a planet on the other side of the Sun, and finally returns back to the Earth along the same path. The total time of the radar echo is
	\begin{equation}\label{eq3_3_10}
		\begin{aligned}
			\Delta T&=2t\left(r_{0}\rightarrow r_{\mathrm{E}}\right)+2t\left(r_{0}\rightarrow r_{\mathrm{p}}\right)\\
			&=\frac{2\left(r_{\mathrm{E}}+r_{\mathrm{p}}\right)}{c}+\frac{4GM}{c^{3}}\ln\frac{4r_{\mathrm{E}}r_{\mathrm{p}}}{r_{0}^{2}}+\frac{4GM}{c^{3}}-\frac{3aGM\pi}{c^{3}r_{0}}\\
			&=\frac{2\left(r_{\mathrm{E}}+r_{\mathrm{p}}\right)}{c}+\Delta T_{\mathrm{GR}}+\Delta T_{a},
		\end{aligned}
	\end{equation}
	where $r_{\mathrm{E}}$ is the Earth-Sun distance, $r_{\mathrm{p}}$ is planet-Sun distance and $r_{0}$ is radar wave-Sun distance when at perihelion.
	\begin{equation}\label{eq3_3_11}
		\Delta T_{\mathrm{GR}}=\frac{4GM}{c^{3}}\ln\frac{4r_{\mathrm{E}}r_{\mathrm{p}}}{r_{0}^{2}}+\frac{4GM}{c^{3}}
	\end{equation}
	is time delay caused by effect of GR. And
	\begin{equation}\label{eq3_3_12}
		\Delta T_{a}=-\frac{3aGM\pi}{c^{3}r_{0}}
	\end{equation} 
	is extra time delay due to non-commutative geometry.
	
	$\Delta T$ is not a directly measurable quantity~\cite{unmeasurable}. In experiments, researchers usually rewrite the term $\frac{4GM}{c^{3}}\ln\frac{4r_{\mathrm{E}}r_{\mathrm{p}}}{r_{0}^{2}}$ in Eq.~\ref{eq3_3_11} as $\frac{2\left(1+\gamma\right)GM}{c^{3}}\ln\frac{4r_{\mathrm{E}}r_{\mathrm{p}}}{r_{0}^{2}}$, and measure the value of $\gamma$~\cite{timedelaygamma}. When $\gamma=1$, it goes back to the predicition of GR.

\begin{figure}[htbp]
		\centering
		\includegraphics[width=1\textwidth]{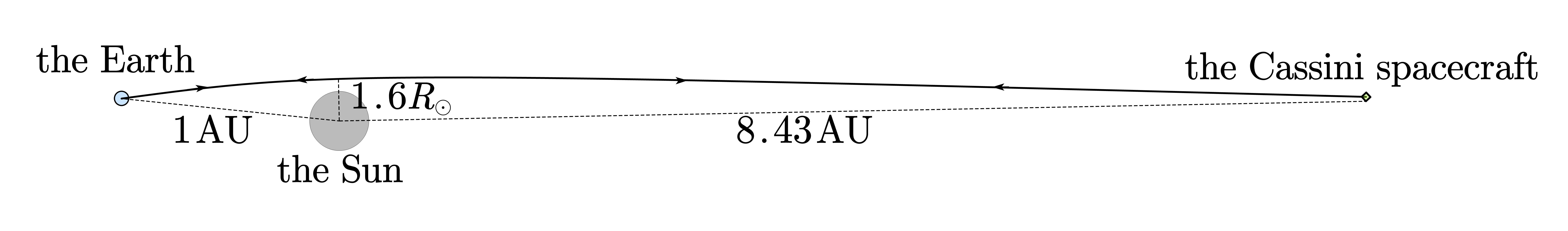}
		\caption{The superior conjunction of the Cassini probe in 2022.}\label{Cassini}
	\end{figure}

	For superior conjunction of the Cassini probe in 2022 (as seen in Fig.~\ref{Cassini}), the spacecraft was at $r_{\mathrm{p}}=8.43$ AU from the Sun. The closest distance of radar wave to the Sun is $r_{0}=1.6R_{\odot}$. The distance of the Earth to the Sun is $r_{\mathrm{E}}=1$ AU. In this experiment, $\gamma$ was measured as $\gamma=1+\left(2.1\pm2.3\right)\times10^{-5}$~\cite{Cassini}. One could further substitute the solar system data~\cite{nasadata} and derive this following constraint
	\begin{equation}\label{eq3_3_13}
		-1.3844\times10^{5}~\mathrm{m}\leq a \leq 6.2925\times10^{3}~\mathrm{m},
	\end{equation}
	or
	\begin{equation}\label{eq3_3_14}
		\Theta \leq 1.9436\times10^{6}~\mathrm{m}^{2}.
	\end{equation}

	\subsection{Gravitational redshift}\label{Sect3_4}
	Considering that an atom emitting a beam of light, which propagates from the surface of the Sun to an observer on the Earth. Since the wavenumber $n=\nu\mathrm{d}\tau$ is a conserved quantity, where $\nu$ represents the frequency of the light and $\mathrm{d}\tau$ denotes the proper time interval, one could have
	\begin{equation}\label{eq3_4_1}
		\nu_{1}\mathrm{d}\tau_{1}=\nu_{2}\mathrm{d}\tau_{2},
	\end{equation}
	where $\nu_{1}$ is the proper frequency of the light wave emitted by a stationary atom on the surface of the Sun, while $\nu_{2}$ is the frequency measured by a stationary observer on the Earth. $\mathrm{d}\tau_{1}$ and $\mathrm{d}\tau_{2}$ are the proper time intervals at the surface of the Sun and at the Earth, respectively. With the relation $\mathrm{d}s^{2}=-c^{2}\mathrm{d}\tau^{2}=-c^{2}f\left(r\right)\mathrm{d}t^{2}$, one have
	\begin{equation}\label{eq3_4_2}
		\frac{\nu_{2}}{\nu_{1}}=\frac{\mathrm{d}\tau_{1}}{\mathrm{d}\tau_{2}}=\sqrt{\frac{f\left(R_{\odot}\right)}{f\left(1\mathrm{AU}\right)}}.
	\end{equation}
	The redshift factor is defined as
	\begin{equation}\label{eq3_4_3}
		z=-\frac{\Delta\nu}{\nu_{1}}=-\frac{\nu_{2}-\nu_{1}}{\nu_{1}}.
	\end{equation}
	The redshift factor is finally obtained as
	\begin{equation}\label{eq3_4_4}
		z=\frac{GM_{\odot}}{c^{2}}\left(\frac{1}{R_{\odot}}-\frac{1}{1\mathrm{AU}}\right)-\frac{aGM_{\odot}}{2c^{2}}\left(\frac{1}{R_{\odot}^{2}}-\frac{1}{1\mathrm{AU}^{2}}\right),
	\end{equation}
	where $\frac{GM_{\odot}}{c^{2}}\left(\frac{1}{R_{\odot}}-\frac{1}{1\mathrm{AU}}\right)$ is the prediction of GR. Due to the Doppler effect, an observer moving away from the wave source can also observe the redshift phenomenon. The Doppler frequency shift formula is given by
	\begin{equation}\label{eq3_4_5}
		\nu_{2}=\nu_{1}\sqrt{\frac{1+\frac{v}{c}}{{1-\frac{v}{c}}}},
	\end{equation}
	where $v$ is the speed of the observer relative to the wave source. If $v>0$, it indicates that the observer is moving toward the wave source, resulting in a blueshift. Conversely, if $v<0$, it indicates that the observer is moving away from the wave source, resulting in a redshift. The redshift factor given by the Doppler effect is
	\begin{equation}\label{eq3_4_6}
		z=-\frac{v}{c}.
	\end{equation}
	Therefore, the redshift factor $z$ could be translated to an equivalent velocity
	\begin{equation}\label{eq3_4_7}
		\left|v\right|=\frac{GM_{\odot}}{c}\left(\frac{1}{R_{\odot}}-\frac{1}{1\mathrm{AU}}\right)-\frac{aGM_{\odot}}{2c}\left(\frac{1}{R_{\odot}^{2}}-\frac{1}{1\mathrm{AU}^{2}}\right).
	\end{equation}
	Regarding the gravitational redshift of the Sun, a precise observational result is the work conducted by J. I. Gonz\'alez Hern\'andez and his team in 2020~\cite{redshift}. They measured data for Fe lines with equivalent widths (EWs) in the range $150<\mathrm{EW}(\mathrm{m}\mathring{\mathrm{A}})<550$, obtaining the result
	\begin{equation}\label{eq3_4_8}
		\left|v\right|=639\pm14~\mathrm{m\cdot s^{-1}}.
	\end{equation}
	Substituting the solar data for calculations~\cite{nasadata}, this observation leads to another constraint on the non-commutative parameter
	\begin{equation}\label{eq3_4_9}
		-4.2973\times10^{7}~\mathrm{m}\leq a \leq 1.8255\times10^{7}~\mathrm{m},
	\end{equation}
	or
	\begin{equation}\label{eq3_4_10}
		\Theta\leq 1.6358\times10^{13}~\mathrm{m^{2}}.		
	\end{equation}
	It should be noted that, since convective motions in the solar photosphere do not affect the iron spectral lines with EW greater than approximately $150\mathrm{m}\mathring{\mathrm{A}}$, the calculations here are based solely on samples with line widths falling within the range $150\mathrm{m}\mathring{\mathrm{A}}<\mathrm{EW}<550\mathrm{m}\mathring{\mathrm{A}}$. However, observations of weak Fe line (with $\mathrm{EW}<180\mathrm{m}\mathring{\mathrm{A}}$) by J. I. Gonz\'alez Hern\'andez's team yielded the result $\left|v\right|=638\pm6~\mathrm{m\cdot s^{-1}}$~\cite{redshift}, which also aligns with the prediction of GR. Moreover, our calculations here do not account for the gravitational effects of the Earth. Since Einstein equations are non-linear, it is not feasible to directly superimpose the metrics generated by the Sun and the Earth for calculation. However, given that the gravitational fields of both the Sun and the Earth are relatively weak, under the weak-field approximation, one could only subtract the contribution generated by the Earth's gravity from the result. However, this effect is so weak (about $\left|v\right|=0.21~\mathrm{m\cdot s^{-1}}$) that it can be entirely neglected.

	\subsection{A brief summary}\label{Sect3_5}
	\begin{table}[h!]
		\centering
		\caption{Constraints on the non-commutative parameter $\Theta$ from solar system tests.}
		\setlength{\tabcolsep}{5.20mm}{
			\begin{tabular}{ccc}
				\hline
				\hline
				Solar tests&Constraints&References of observation\\
				\hline
				Mercury precession&$\Theta\leq0.067579~\mathrm{m}^{2}$&Ref.~\cite{Mercury}\\
				Light deflection&$\Theta\leq6.8481\times10^{8}~\mathrm{m^{2}}$&Ref.~\cite{bending}\\
				Time delay of light&$\Theta \leq 1.9436\times10^{6}~\mathrm{m}^{2}$&Ref.~\cite{Cassini}\\
				Gravitational redshift&$\Theta\leq 1.6358\times10^{13}~\mathrm{m^{2}}$&Ref.~\cite{redshift}\\
				
				\hline
				\hline
		\end{tabular}}
		\label{tab1}
	\end{table}
	From Eqs.~\ref{eq3_1_24}, \ref{eq3_2_15}, \ref{eq3_3_12} and \ref{eq3_4_7}, one can observe that non-commutative geometry exerts a weakening effect on the gravitational interactions. Furthermore, the influence of non-commutative geometry on the geodesic motion of time-like particles is much more pronounced than its effect on the motion of photons. Tab.~\ref{tab1} is a summary of parameter constraints given by the solar tests. It is evident that the value of the non-commutative parameter inferred from solar system experiments should lie within the range of $\Theta\leq0.067579~\mathrm{m}^{2}$. The scale of the Planck length clearly falls within this range. In comparison, the range of $\Theta$ provided by the experiments appear to be excessively large. However, considering that the motions discussed in this section are located outside the solar surface, even for the result with the maximum width among the four datasets, there remains $\sqrt{1.6358\times10^{13}}~\mathrm{m}/R_{\odot}=5.8\times10^{-3}\ll1$. Therefore, employing Eq.~\ref{eq2_6} for computations is deemed sufficiently safe. And, on astronomical scales, these constraints already represent relatively precise limits.

	\section{Conclusion and outlook}\label{Sect4}
	We conducted detailed calculations of the four classical tests of GR in Schwarzschild space-time with a Lorentzian distribution, including planetary orbital precession, light deflection, radar wave delay, and gravitational redshift effects. The study reveals that the non-commutative parameter has a markedly stronger impact on the motion of time-like particles compared to its influence on light rays. With precise experimental observations, the parameter is ultimately constrained within the range $\Theta\leq0.067579~\mathrm{m}^{2}$, which is given by the observations of Mercury's precession. This result is consistent with the view that $\sqrt{\Theta}$ is on the order of the Planck length $\ell_{p}$. Moreover, this range exceeds the Planck scale by several orders of magnitude. However, considering that our study focuses on large-scale gravitational experiments within the solar system, it is expected that the derived parameter range will indeed surpass the Planck length. Within the scale of astronomical measurements, this result presents a relatively stringent constraint.
	
	It should be noted that this paper only considered the low-order correction of the non-commutative parameter. Taking into account higher-order corrections of non-commutative geometry would help provide more accurate understanding of its implications in gravitational theory. Additionally, because the external space-time of the Sun is equivalent to the gravitational field produced by a point mass, the direct application of the Lorentzian distribution modified by point particles is reasonable. However, the Sun itself is a celestial body with a certain volume and a complex mass distribution. If we model the Sun as a perfect fluid, studying the gravitational field generated by a perfect fluid in non-commutative geometry would hold greater practical significance. Besides, when different distributions (such as Gaussian distributions) are considered, the upper limit of the non-commutative parameter $\Theta$ will also change. This paper only discusses the Schwarzschild space-time modified by Lorentz corrections. Moreover, considering the effect of gravitational radiation, the orbits of objects in motion around the gravitational center will continuously shrink. A further discussion of the effects of shrinking orbits will provide more precise constraints on the parameters. However, considering that gravitational effect within the solar system is not significant, this effect is expected to be very weak. Furthermore, it should be noted that in~\cite{Hyperbolic}, researcher has calculated the general relativistic effects on unbound hyperbolic trajectories, where the method used can be generalized to any modified gravity models. Applying this method to non-commutative geometry and further exploring the effects of Lorentzian or Gaussian distributions on hyperbolic trajectories would constitute a novel topic of research.

	\section*{Conflicts of interest}
	The authors declare that there are no conflicts of interest regarding the publication of this paper.
	
	\section*{Acknowledgments}
	We want to thank the National Natural Science Foundation of China (Grant No. 11571342) for supporting us on this work.
	
	\section*{Data availability statement}
	The data generated in this study are available on a reasonable requirement.
	
	\bibliographystyle{unsrt}
	\bibliography{paper}
\end{document}